# SWift -A SignWriting editor to bridge between deaf world and e-learning

Claudia Savina Bianchini – Fabrizio Borgia – Maria De Marsico

*Abstract* — SWift (SignWriting improved fast transcriber) is an advanced editor for SignWriting (SW). At present, SW is a promising alternative to provide documents in an easy-to-grasp written form of (any) Sign Language, the gestural way of communication which is widely adopted by the deaf community. SWift was developed SW users, either deaf or not, to support collaboration and exchange of ideas. The application allows composing and saving desired signs using elementary components, called *glyphs*. The procedure that was devised guides and simplifies the editing process. SWift aims at breaking the "electronic" barriers that keep the deaf community away from ICT in general, and from e-learning in particular. The editor can be contained in a pluggable module; therefore, it can be integrated everywhere the use of SW is an advisable alternative to written "verbal" language, which often hinders information grasping by deaf users.

## I. INTRODUCTION

In 2006, according to the World Health Organization, 278 million people worldwide were deaf or had hearing difficulties. Deafs experience many difficulties in communicating with hearing people. Moreover, many of them, even if they know a verbal language, prefer to communicate through a visual-gestural form defining national Sign Languages (SLs). The perceptual basis of such languages helps expressing one's own thoughts in ways different from written/spoken languages that rely on the phonetic experience which deafs lack. The structure of a SL is deeply different from the sequential frame of a written/spoken language, due to the use of multi-linear (both spatial and temporal) relationships among gestures and their components (manual and non-manual). In approaching the digital world, deafs must overcome barriers similar to those faced in their everyday life. "Regional variations of sign language form a group of relatively under-represented language minorities in the digital world. Thus, members of the deaf community are usually confronted with websites that do not include their preferred tongue, and this may cause accessibility barriers. To ensure the social integration of the deaf community, sign languages should be properly incorporated into the Information Technology" [01]. The lack of specific guidelines does not help to overcome the digital divide. The only attempts to address general accessibility issues have been made by the World Wide Web Consortium (W3C), with the Web Content Accessibility Guidelines (WCAG) document. However, the most addressed disability is blindness, while there is scarce understanding of the problems encountered by deaf people in acquiring and using verbal languages (see for example [02]). A written transcription of audio content is considered enough to support such users. In fact, WCAG1.0 guidelines (1999) deal mostly with problems related to labeling and transcription of audio content, leaving out alternatives related to SLs. WCAG2.0 (2008) better addresses such issues: e.g., the success criterion 1.2.6, whose satisfaction is necessary to get the highest level of compliance (AAA), states that "Sign language interpretation is provided for all prerecorded audio content in the form of synchronous media types." On the other hand, the lack of a written form for SLs still hinders their wider use. Moreover, since information is conveyed in a multi-linear way, through gaze, hands, facial expression, head and shoulders, it is unfeasible to transcribe them by the written form of a VL. The difficulty to provide information in a form easily comprehensible by deaf users pertains to all the resources available for distance learning. Different disabilities call for specific communication strategies. The appropriate use of technology represents a unique opportunity for learners with disabilities to participate in activities which are obvious and easy for the majority; on the other hand, if deaf learners cannot take advantage of them, they might be further penalized by the lack of appropriate learning contents. The SWift project is mired at supporting the use of SignWriting, which seems, at present, a very promising solution for SLs transcription.

## II. SIGNWRITING FOR LEARNING RESOURCES

*A. Vygotsky and accessibility*

Vygotsky's theory of proximal development [03] is related to two crucial aspects of modern educational strategies. On one side, the concept of Zone of Proximal Development (ZPD) calls for personalized learning paths and scaffolding.

On the other side, the social aspect of learning is considered of paramount importance, taking to socio-constructivism and situated learning. Communication and collaboration imply exchange of data and experience, and today more than ever this is made possible by ICT technologies, especially Web2.0. However, users with disabilities risk facing a renewed version of the digital divide. If individual ZPDs may call for similar contents, these must be deployed according to parallel strategies. The design of accessible e-learning instruments is often targeted to actively involve people with disabilities in the socially relevant phenomenon of distance learning. Different "wrappers" should be provided to impaired people, and this is a demanding issue that deserves particular attention and effort. A careful design can identify the critical conceptual building blocks and specify the modes of interaction provided to different kinds of users [04]. Unfortunately, there is often a lack of suitable tools. SignWriting is a promising medium to support deaf users.

*B. Means to convey contents for deaf users*

An often exploited solution to convey on the Web meanings for deaf people is to integrate information with videos showing a person using SL to express them. This strategy is often used to create whole sites by exclusively using SL, for both content and navigation paths. A basis for this approach is Hypervideo technology [05]. Links are inserted in cutscenes of a video in the form of text or static images. Based on this technology, Fels *et al*. ([06]) developed Signlinking. Each Signlink is a time slice within the video clip. When the video reaches a Signlink, a link indicator is displayed. Though aesthetically and technically attractive, a similar solution presents two main drawbacks: it is both technically and time engaging to set up recording of a high number of clips to transmit the complete information; in addition, video on the web often suffers from bandwidth limitations. A "written" form for SLs would address the above problems. The first notations devised to "write" the content of a signed sequence represented the manual component of SLs according to the place where it is produced, to the configuration of hands, and to their movement. Further parameters were later identified in palm orientation, gaze, and in some observed regularities. However, such notations required annotations in verbal language.

SignWriting (SW) was devised by the choreographer Valerie Sutton [07], and is a completely visual notation. It is therefore a promising candidate as written companion to SLs. Individual signs result from the composition of highly iconic basic elements called *glyphs*. These components identify the sign, namely the configurations of the body parts involved, contact points, dynamics, etc. Figure 1 shows an example of the prototypical SW notation. Glyphs make up a sort of rich alphabet, able to represent signs in any SL in a bidimensional space. Let us note that a universal Sign Language does not exist: as similar gestures have completely different meanings in different countries and cultures, national SLs present often deep differences.

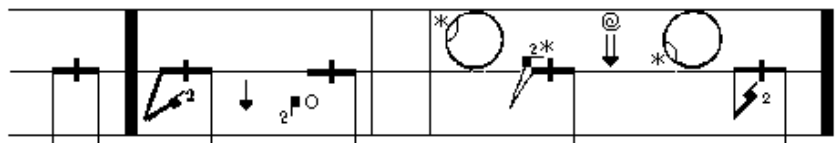

Figure 1. Example of SignWriting notation (1974).

*C. The SWORD frame work*

The overall goal of our work is to make SignWriting effectively exploitable as a communication mean, and a suitable learning support for deaf people. Project SWORD (SignWriting Oriented Resources for Deafness) aims at supporting this written form for SLs in many media formats.

Along this way, a first web-based tool is SWift (SignWriting improved fast transcriber), that we developed with the advice of experts and deaf researchers. It currently allows to easily compose single signs. We started its design considering the features of a Web application of similar use, namely SignMaker; the next step will be to expand the tool to accommodate entire documents in SignWriting. Moreover, we want to build a software environment that can also handle signs coming from other sources, such as hand written documents in SignWriting or video clips of signing people, and translate them in electronic form.

III. SWIFT WEB APPLICATION

*A. SignMaker, the starting point*

The aim of Swift is electronic editing in SW. With the same aim, the team of Valerie Sutton has created and developed SignMaker, a web-based application to write and save signs. Figure 2 shows its home screen.

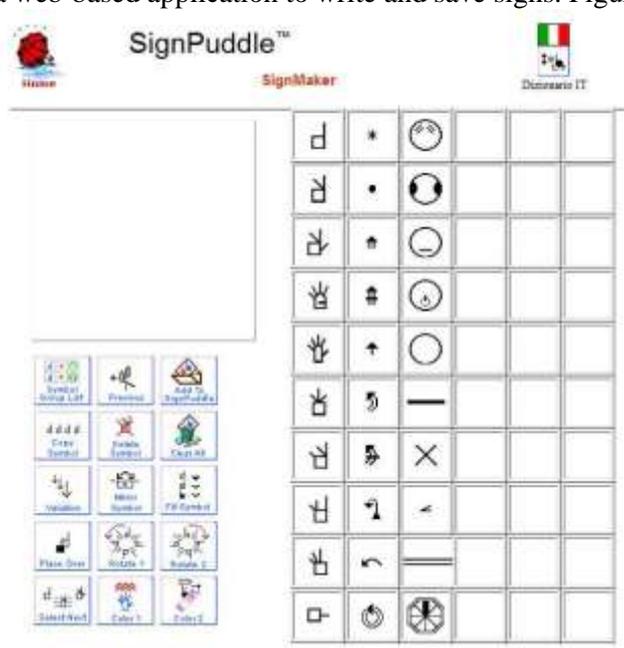

Figure 2. SignMaker's home screen

The upper left area is the *Sign Display*, a whiteboard where the sign is composed, one glyph at a time. The right area, the *Glyph Menu*, is organized in a tree of submenus allowing the choice of the glyph to drag in the Display. Glyphs can be freely positioned, consistently with SW lack of constraints. The lower left area contains the *Toolbox*, with buttons triggering tasks like editing selected glyphs, resetting the Sign Display, etc. Once inside the Display, the glyphs remain draggable, and become selectable one at a time, to be handled using the functions in the Toolbox.

There are some critical aspects for improving. The glyphs in the Sign Display must be selected one at a time, making it impossible to edit more of them simultaneously (e.g. deleting or rotating). Moreover, the functions for managing the Sign Display are far from their target, and hidden among other functions in the Toolbox.

The Glyph Menu is implemented as a tree menu: from the "root" (the home, the only menu from which one can reach all the others) the user navigates towards a "leaf" menu. Any glyph contained in any menu (not just those in the "leaf" menus) is draggable on the Sign Display. The SignMaker offers basic navigation functions: a button to return to the previous menu and one to return to the home of the Glyph Menu. During sign editing, most of the time is spent in navigating through the Glyph Menu, whose structure and provided interaction scarcely guide the user. Moreover, as for the Sign Display, the associated navigation functions are mingled in the Toolbox. The most serious problem is the spatial arrangement of the navigation menu. The groups of glyphs of different type are presented next to each other, so that users, with their knowledge of SW, are completely responsible for choosing the correct search path.

The Toolbox is the area where more space for improvements has been detected. Navigation buttons for the Glyph Menu are placed aside and mingle with those (the majority) for handling the selected glyphs in the Sign Display, and with those for handling the Sign Display itself. A good rule in human-computer interaction is to use icons self-expressing their function, without the need for text. In this case the rule cannot be overlooked, since deaf users may have difficulties in understanding text. In addition, the text on the buttons is in English, and there's no way of choosing another language. Most icons can be fully understood only by the small community of signing people with a good knowledge of SW.

*B. SWift interface*

Figure 3 shows the home screen of SWift interface. It minimizes the use of text labels and presents a collection of colorful and familiar icons. The goal is to avoid immediately confusing the users with a large amount of information, and to relieve them from the burden of learning the program.

The application has been designed in strict collaboration with deaf users, according to the core principles of User-Centered ([08]) and Contextual ([09]) design. In particular, we worked with the Team of ISTC-CNR, which includes deaf researchers, who are a true sample of the main target users. As an example, the almost complete absence of textual labels derives from precise needs expressed by them. The redesign of the graphics has been joined with a complete redesign of the logic part of the application. Three areas of SWift interface have the same names and functions as in SignMaker, while the Hint Panel is an innovation.

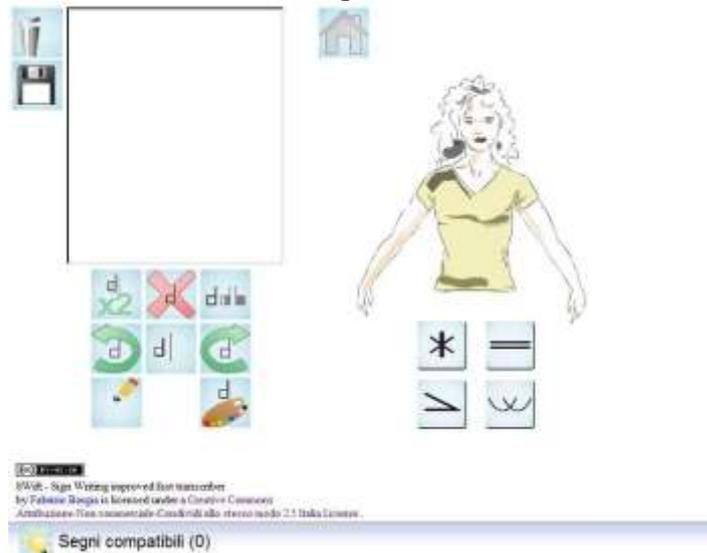

Figure 3. Home screen of SWift.

**Sign Display**: the upper right area of Figure 3 is a whiteboard on which the sign is composed, one glyph at a time; a pair of buttons just on the left is used for its management.

**Toolbox**: the lower right area of Figure 3 contains buttons to operate on the glyphs which are already in the Sign Display.

**Glyph Menu**: the left part of Figure 3 is a set of graphic elements that allow the user to choose the glyphs to drag in the Sign Display; these elements support an intuitive search.

**Hint Panel**: the lower band in Figure 3 shows a series of glyphs that belong to the anatomical area currently explored, and are compatible with those in the Sign Display; compatible means that are found together with a significant frequency in the signs stored in the database; when the panel is minimized, a textual label shows the number of compatible glyphs found, otherwise the usual drag & drop allows to enter one of such glyphs in the Sign Display.

Glyphs in the Sign Display remain draggable and become selectable to be edited through the functions in the Toolbox: differently from SignMaker, it is possible to select multiple glyphs at once, thus allowing to save time for handling them in the same way. The two buttons on the left of the Sign Display allow to clear the whole content of the Sign Display, or to save the composed sign. The save popup menu allows to choose among three formats: text, image, and the format for SWift database. It was not possible to avoid the use of the text in this popup. For instance, there is no icon to represent the png format, so it was necessary to write this using a text label. Insertion in SWift database implies storing the sign with the detailed list of its component glyphs. This action is the base for the statistical computation of compatibility.

Buttons in the Toolbox deserved particular attention. Each icon is 'animated' by a short sequence triggered by mouseover, mimicking the performed action. Using an animation rather than a clip with a signed sequence requires less space and time resources. An example is in Figure 4, showing screenshots from anti-clockwise rotation button.

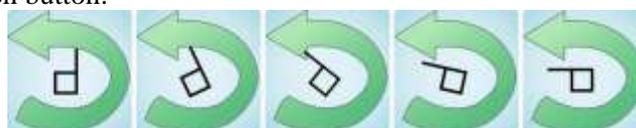

Figure 4. Anti-clockwise rotation button animation.

Most design and optimization were focused on the interaction with the Glyph Menu. Making this interaction more efficient reduces the search time, which in turn determines composition time of each sign. The stylized human figure, named "Puppet", which is initially presented, and the buttons beneath, presents the first important choice. The areas corresponding to relevant anatomical elements (head, shoulders, hands, arms) will highlight on mouseover, suggesting the user the possibility to click and choose one. The user can initially choose one anatomic area of the Puppet, or one of the buttons which represent items such as punctuation and contacts. This choice takes to the appropriate dedicated search menu. As an example, clicking on the hands, the user will access the search menu allowing to choose related configurations and motions (see Figure 5). After the first user's choice, the Puppet and the buttons beneath are reduced and shifted to the left, and remain clickable, to form a navigation menu together with the button to return to the Glyph Menu's home screen. This allows navigating from one area to another without having to re-pass from the home screen, unless using the ad-hoc button. A red square appears around the selected area to remind the performed choice, like breadcrumbs. In the central part of the menu, a label and an icon explain the user what kind of glyphs are available using the group of boxes beneath. Each of such blue rectangles, named "Choose Boxes", identifies possible (incompatible) choices for a certain glyph feature. The user can search for the desired glyph according to one or more features at the same time, by selecting each feature value from the appropriate box. No predefined order constrains the sequence of choices. For example, a user might choose the glyph rotation first, while another could choose the handedness (left, right), or the number of fingers.

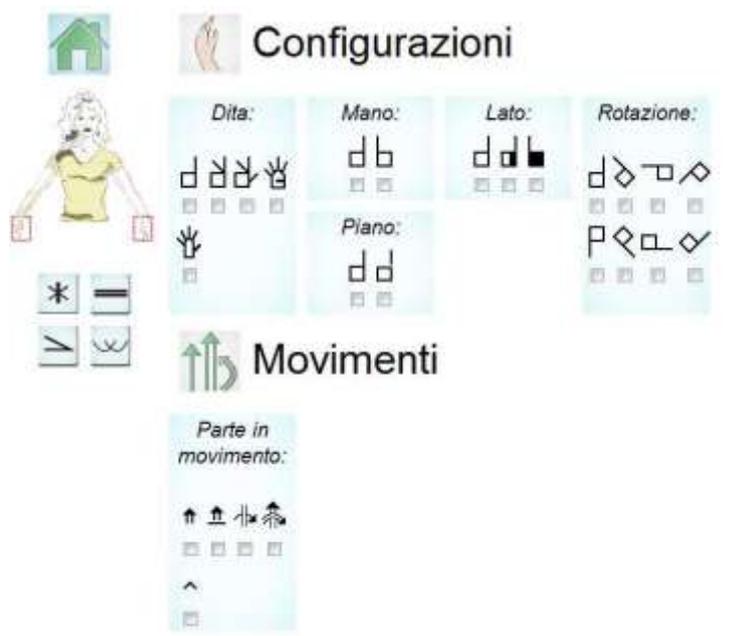

Figure 5. Glyph Menu -Search menu for the hands.

Glyphs corresponding to the current selected criteria are placed into a scrollable panel and are immediately draggable into the Sign Display. Once either a choice is made in any of the Boxes, or a previous one is changed or simply canceled (undo), the system updates the panel by respectively selecting the correct subset from the set of glyphs currently displayed, or by modifying the current set according to the changed feature value. In very lucky cases the user can find the desired glyph by making only one choice. More often, about 3 features must be defined to find the searched glyph.

Some areas, e.g. hands, contain a large number of glyphs, so more set of Boxes may be found: in Figure 5 we see that the user can choose to search according to movement or to configuration, both related to hands. Once the first choice is made, the unnecessary Boxes leave room for the search results. For instance, choosing to search a glyph representing a configuration (e.g. the number of fingers) replaces the movement Boxes with the search results.

## IV. USABILITY EVALUATION

Deaf users cannot actually "think aloud". They can rather express their thoughts through their SL, and often demonstrate an high variability in their face expression. For these reasons, Roberts and Fels ([10]) suggest the setting shown in Figure 6. CAM1 records a rear view of the participant, the computer screen, and the interpreter. CAM2 records the front view of the participant, and the investigator. Using two cameras requires analyzing and synchronizing data from two recordings. Furthermore, it is difficult to maintain a synoptic view of everything that happens in the environment at any time. Therefore, we adopted a system using a single camera, using a projector to send the computer screen image on the wall, so that a single camera records everything worth of attention. The test was structured in three phases: in the *welcome time*, the participant is briefed by a screen containing a signed video and its transcription; during the *sign-aloud test*, the participant is asked to perform a list of tasks using SWift, and to sign anything that comes to her/his mind; a final *usability questionnaire* is adapted from QUIS ([11]).

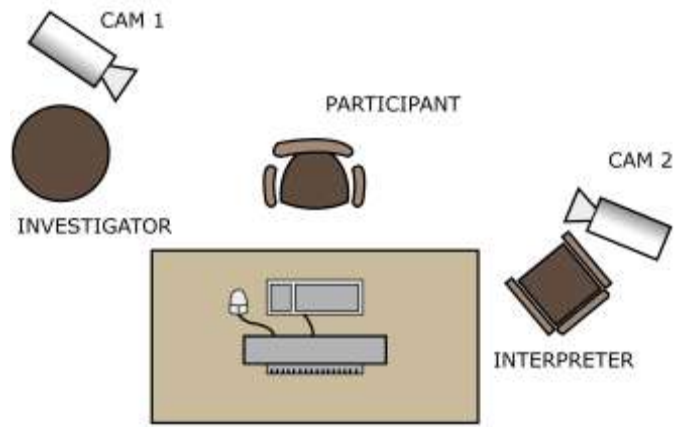

Figure 6. Spatial setting suggested in ([10])

The central phase deserved special attention. Users need reminders for which task is being performed, what needs to be done, etc. The task list should be available in both Verbal Language (VL) and SL. Besides having the list in electronic form, we can either involve an SL interpreter or not, and with a varying role. We chose to involve the interpreter because the possibility of interaction with the participant increases the correct understanding of the tasks. In particular, our interpreter always provided a task translation in SL at the beginning of each task. Required tasks included both basic actions, such as inserting a random glyph or looking for a particular one, and complex ones, such as composing an assigned sign. The questionnaire in the third phase was designed adapting the QUIS usability questionnaire to the specific application and to the needs of deaf users, to stimulate participants to express their opinions. Each (simplified) written question has a corresponding SL clip.

We ran a preliminary test session with ten deaf users. Results were very encouraging, and will help to solve some problems. We plan to perform a more extensive evaluation in the near future. It is to consider that the typical problems of user recruitment for tests are amplified by the peculiarities of the deaf condition, especially since knowledge of SignWriting is still limited, and by the need of a SL interpreter. We are investigating a modified version of the tests to be run online.

The low number of errors made by participants in the use of interface buttons confirmed the quality of most of our choices. However, as for the navigation, the lack of any glyph in the home screen was often misleading. This may be due to their former confidence with SignMaker interface. When arriving to a specific set of Choose Boxes, many deaf users made only one choice, and then browsed the whole resulting set (about 400 elements). We are investigating a way to make clearer the possibility of choosing one option per box. At the end of the test session, most users expressed appreciation for the test modalities, in particular for the final questionnaire.

Froma a purely quantitative point of view, insertion times (average over 15 glyphs) using SM (01:15:10) and Swift (00:57:30) confirm a better interaction flow.

## V. Conclusions and future work

Our goal is to achieve a comprehensive SW handling. However, processing of handwritten SW documents and of videos requires a substantial processing. Hand-written documents need scanning and storing in electronic form, and individual signs must be identified. We plan to implement a sort of OCR (Optical Character Recognition) starting from the state of the art on the ideographs recognition. The last step of the project is the processing of video sequences of signing subjects, in order to identify the signs, and then transcribe them into SignWriting. The pattern recognition procedures involved in this case are much more complex, therefore we plan to implement a "user assisted" procedure to simplify the task.

## Acknowledgment

This work was supported by Italian MIUR under the FIRB project "VISEL – E-learning, deafness, written language: a bridge of letters and signs towards knowledge society". We thank the deaf and hearing researchers at IST-CNR of Rome for their support and suggestions during the design and evaluation of SWift. In memory of Elena Antinoro Pizzuto.

## References

[01] I. Fajardo, M. Vigo, and L. Salmeron, "Technology for supporting web information search and learning in sign language". *Interacting with Computers*, vol. 21, no. 4, pp. 243–256, 2009.

[02] C.A. Perfetti and R. Sandak, "Reading optimally builds on spoken language: implications for deaf readers." Journal of Deaf Studies and Deaf Education, vol. 5, pp. 32–50, 2000.

[03] L.S. Vygotsky, *Mind in society: the development of higher psychological processes*. M. Cole, V. John-Steiner, S. Scribner, and E. Souberman eds., Harvard University Press, 1978

[04] M. De Marsico, S. Kimani, V. Mirabella, K. Norman, and T. Catarci, "A proposal toward the development of accessible e-learning content by human involvement". *Universal Access in Information Society*, vol. 5, no. 2, pp. 150–169, 2006.

[05] M. Debevc, R. Safaric, and M. Golob, "Hypervideo application on an experimental control system as an approach to education". *Computer Applications in Engineering Education*, vol. 16, no. 1, pp. 31–44, 2008.

[06] D.I. Fels, A. Richards, J. Hardman, and D.G. Lee, "Sign language web pages". *American Annals of the Deaf*, vol. 151, no. 4, pp. 423–433, 2006.

[07] V. Sutton, "A way to analyze American sign language & any other sign language without translation into any spoken language". in *National Symposium on Sign Language Research and Teaching*, 1980.

[08] D.A. Norman, and S.W. Draper, *User centered system design: new perspectives on human-computer interaction*. CRC Press, 1986.

[09] D. Wixon, K. Holtzblatt, and S. Knox, "Contextual design: an emergent view of system design". in *Proceedings CHI'90*. ACM, pp. 329–336, 1990.

[10] V.L. Roberts, and D.I. Fels, "Methods for inclusion: employing think aloud protocols in software usability studies with individuals who are deaf". *International Journal of Human-Computer Studies*, vol. 64, no. 6, pp. 489–501, 2006.

[11] L. Slaughter, K.L. Norman, and B. Shneiderman, "Assessing users' subjective satisfaction with the information system for youth services (isys)," in *VA Tech Proceedings of Third Annual Mid-Atlantic Human Factors Conference*, pp. 164–170, 1995.